# Heterointerface effects on lithium-induced phase transitions in intercalated MoS$_2$

Sajad Yazdani[1,2,†], Joshua V. Pondick[1,2,†], Aakash Kumar[1,2], Milad Yarali[1,2], John M. Woods[1,2], David J. Hynek[1,2], Diana Y. Qiu[1,2], Judy J. Cha[1,2*]

[1]Department of Mechanical Engineering and Materials Science, Yale University, New Haven, CT 06511, USA.
[2]Energy Sciences Institute, Yale West Campus, West Haven, CT 06516, USA.
[†]These authors contributed equally.
[*]Correspondence to: judy.cha@yale.edu

**Abstract**

The intercalation-induced phase transition of $MoS_2$ from the semiconducting 2H to the semimetallic 1T' phase has been studied in detail for nearly a decade; however, the effects of a heterointerface between $MoS_2$ and other two-dimensional (2D) crystals on the phase transition have largely been overlooked. Here, *ab initio* calculations show that intercalating Li at a $MoS_2$-hexagonal boron nitride (*h*BN) interface stabilizes the 1T phase over the 2H phase of $MoS_2$ by ~ 100 mJ m$^{-2}$, suggesting that encapsulating $MoS_2$ with *h*BN may lower the electrochemical energy needed for the intercalation-induced phase transition. However, *in situ* Raman spectroscopy of *h*BN-$MoS_2$-*h*BN heterostructures during electrochemical intercalation of Li$^+$ shows that the phase transition occurs at the same applied voltage for the heterostructure as for bare $MoS_2$. We hypothesize that the predicted thermodynamic stabilization of the 1T'-$MoS_2$-*h*BN interface is counteracted by an energy barrier to the phase transition imposed by the steric hindrance of the heterointerface. The phase transition occurs at lower applied voltages upon heating the heterostructure, which supports our hypothesis. Our study highlights that interfacial effects of 2D heterostructures can go beyond modulating electrical properties and can modify electrochemical and phase transition behaviors.

**Main Text**

Van der Waals heterostructures comprised of different two-dimensional (2D) materials[1,2] can exhibit novel electronic and optical properties[3–5], in which the effect of heterointerfaces is important. Heterointerfaces may also play a central role in determining electrochemical properties and phase transitions of 2D materials. For $MoS_2$, intercalation of alkali metal ions such as Li$^+$ into the van der Waals gaps between the layers can induce a phase transition from the semiconducting trigonal prismatic 2H phase to the semimetallic octahedral 1T phase[6–8], followed by an additional phase transition to the distorted 1T' phase, through electron doping by the intercalants[9–12]. The intercalation-induced phase transition has proven important for energy applications[13–17], and the thermodynamics of the electrochemical intercalation of Li$^+$ in MoS$_2$ has been well-characterized [18–20]. Despite the extensive investigation and many applications of the intercalation-induced phase transition, the effect of a heterointerface on this process has largely been overlooked.

Herein, we investigate heterointerface effects on the intercalation-induced phase transition of MoS$_2$ by interfacing MoS$_2$ with hexagonal boron nitride (*h*BN). First, *ab initio* calculations indicate that the 1T-MoS$_2$-*h*BN interface is thermodynamically more stable than the 2H-MoS$_2$-*h*BN interface when Li is inserted at the interface, suggesting that interfacing MoS$_2$ with *h*BN could lower the thermodynamic energy barrier for the phase transition. Second, we performed electrochemical lithium intercalation on *h*BN-MoS$_2$-*h*BN heterostructures using electrochemical microreactors[20–26] and studied the intercalation-induced phase change via Raman spectroscopy, *in situ*. In contrast to the *ab initio* results, we observe that the phase transition for *h*BN-MoS$_2$-*h*BN heterostructures occurs at the same applied electrochemical voltage as bare MoS$_2$ irrespective of the Li$^+$ intercalation pathways. Thus, we hypothesize that the phase transition in Li-intercalated *h*BN-MoS$_2$-*h*BN heterostructures is influenced by steric hindrance imposed by the heterointerface, which would counteract the predicted stabilization of the 1T-MoS$_2$-*h*BN interface over the 2H-MoS$_2$-*h*BN interface. *In situ* heating of the heterostructure during intercalation allowed for the phase transition to occur at a lower applied electrochemical voltage, supporting the hypothesis that the heterointerface imposes a barrier to intercalation. Lastly, we demonstrate that the basal planes of commercial *h*BN are ionically conductive for Li$^+$.

**Results**

***Ab initio* calculations of the MoS$_2$-*h*BN interface with Li$^+$ intercalation**

We performed density functional theory (DFT) calculations to investigate the energetics of the 2H to 1T phase transition of MoS$_2$ in the presence and absence of an *h*BN interface (see Methods). We note that the high symmetry 1T structure was used instead of the distorted 1T' structure to reduce computational cost and because 1T is an intermediate state in the 2H to 1T' phase transition. In the absence of *h*BN, the 2H to 1T phase transition in MoS$_2$ is thermodynamically driven, with the 1T phase becoming more favorable (lower in energy) than the 2H phase when bulk MoS$_2$ is doped above -1.3 electrons per monolayer formula unit (f.u.) (Fig. 1a), consistent with previous calculations[27,28]. In the presence of an *h*BN interface, in which monolayer MoS$_2$ is encapsulated by 6-layer *h*BN (Fig. 1b), the binding energy of the *h*BN-MoS$_2$ interface is lower for 1T-MoS$_2$ than for 2H-MoS$_2$ by ~100 mJ m$^{-2}$ when a Li atom is intercalated in the van der Waals gap between MoS$_2$ and *h*BN (Table 1 and see Methods for the heterostructure supercell and details of relaxation). The significant decrease in the interfacial energy for 1T-MoS$_2$-*h*BN over 2H-MoS$_2$-*h*BN with Li intercalated at the interface suggests that the nucleation barrier for the intercalation-induced phase transition would be lower for *h*BN-MoS$_2$-*h*BN heterostructures than for bare MoS$_2$.

We also considered in-plane strain due to the lattice mismatch between MoS$_2$ and *h*BN, which may influence the calculated stability of the 2H- and 1T-MoS$_2$-*h*BN interface. To create a commensurate 3x3 MoS$_2$-4x4 *h*BN supercell, we applied an in-plane strain of roughly 4% to *h*BN, but this strain can be reduced to about 2% by utilizing a larger 4x4 MoS$_2$-5x5 *h*BN supercell. However, the calculated binding energies of the two interfaces did not change in these two strain states (Supplementary Table 1), so we conclude that the in-plane strain on *h*BN did not have an effect on the increased stability of the intercalated 1T-$MoS_2$-*h*BN interface.

Intercalated Li may donate charge to *h*BN at the $MoS_2$-*h*BN interface, which may reduce the effective doping of $MoS_2$. To quantify this doping, we calculated the amount of charge transfer from an intercalated Li atom to the adjacent $MoS_2$ and *h*BN layers. Figure 1c shows the isosurface of the charge density with a Li atom in the van der Waals gap of 2H-$MoS_2$-*h*BN, which shows that the majority of the charge is donated to the three S atoms that are the nearest neighbor of the Li. Some charge is also donated to the N atoms of *h*BN closest to Li (Supplementary Fig. 1). Bader charge analysis[29] shows that the total charge donated to the three S atoms is -0.97 electron for a freestanding monolayer of 2H-$MoS_2$ and is reduced to -0.87 electron in the presence of *h*BN with about -0.1 electron donated to N atoms of *h*BN (Fig. 1d). Notably, the amount of charge transfer falls off rapidly with increasing distance from the Li, with > 90% of the charge transfer confined within the ring of nearest neighbor atoms (Supplementary Fig. 1). Thus, the intercalated Li at the $MoS_2$-*h*BN interface still dopes $MoS_2$ strongly and contributes to the overall electron doping required for the phase transition of $MoS_2$. We therefore predict that the critical electron density and, by extension, lithium concentration required to induce the phase transition, would either be similar or even lowered by interfacing $MoS_2$ with *h*BN, given the lower binding energy of 1T-$MoS_2$-*h*BN interface.

**Electrochemical intercalation of *h*BN-$MoS_2$-*h*BN heterostructures**

Motivated by the theoretical prediction, we performed electrochemical lithium intercalation on *h*BN-$MoS_2$-*h*BN heterostructures in which several-layer-thick $MoS_2$ flakes were sandwiched between exfoliated flakes of hBN (see Methods and Fig. 2a). The hBN-MoS$_2$-hBN heterostructure devices were integrated into electrochemical microreactors[20–24] and intercalation was controlled potentiostatically by sweeping the electrochemical potential (V$_{EC}$) between the hBN-MoS$_2$-hBN working electrode and a lithium reference/counter electrode. Cr/Au metal contacts directly deposited onto the top surface of MoS$_2$ in the heterostructures served as the working electrode, while lithium metal pressed onto copper foil served as the reference/counter electrode (Fig. 2b). As V$_{EC}$ is lowered from the open circuit voltage (O.C.), Li$^+$ ions diffuse through the electrolyte and intercalate into the interlayer gaps of the heterostructure. We note that while multilayer MoS$_2$ flakes are used in experiments unlike the monolayer MoS$_2$ used in calculations, the computational predictions remain valid as they examine the binding energy of the MoS$_2$-hBN interface for 1T- and 2H-MoS$_2$, which we determined to be independent of increasing the thickness of MoS$_2$ up to two layers.

Intercalation dynamics of hBN-MoS$_2$-hBN heterostructures were compared to those of bare MoS$_2$ by sequentially lowering V$_{EC}$ from O.C. (~2.5 V vs. Li/Li$^+$) to 0.2 V vs. Li/Li$^+$ at a decrement of 0.2 V. At each potential, V$_{EC}$ was held constant for several minutes to acquire Raman spectra *in situ*, after which the samples were allowed to recover to O.C. before lowering V$_{EC}$ to the next potential. *In situ* optical microscopy of a heterostructure during intercalation revealed that MoS$_2$ turned dark at 1.0 V vs. Li/Li$^+$ (Fig. 2c), and bare MoS$_2$ also darkened at 1.0 V vs. Li/Li$^+$ (Fig. 2d). The color change may indicate a phase transition to the 1T' phase[21], and *in situ* Raman spectroscopy confirmed the 2H-1T' phase transition of MoS$_2$ in the heterostructure at 1.0 V vs. Li/Li$^+$ and also in bare MoS$_2$ at 1.0 V vs. Li/Li$^+$ (Fig. 2e). The phase transition from 2H- to 1T'-MoS$_2$ is clearly observed via the suppression of the characteristic E$_{2g}$ and A$_{1g}$ Raman modes of 2H-MoS$_2$ and the concurrent growth of the J$_1$ and J$_2$ modes of 1T'-MoS$_2$[30,31]. We note that, due to the low intensity of $h$BN Raman modes (Supplementary Fig. 2), we were unable to determine what effect lithium has on the structure of $h$BN. Supplementary Fig. 3 shows the full intercalation range of the heterostructure, including the transition to the amorphous state, which occurs at 0.4 V vs. Li/Li$^+$, the same as in the bare MoS$_2$ case (Supplementary Fig. 4). Therefore, our key finding is that while encapsulating MoS$_2$ with $h$BN should theoretically significantly increase the thermodynamic stability of 1T'-MoS$_2$, it has no observable effect on the electrochemical voltage required to induce the 2H-1T' phase transition of lithium-intercalated $h$BN-MoS$_2$-$h$BN.

**Lithium diffusion pathways in $h$BN-MoS$_2$-$h$BN heterostructures**

One possible explanation for the lack of an observable difference between bare MoS$_2$ and $h$BN-MoS$_2$-$h$BN heterostructure is an increased Li$^+$ diffusion pathway in the heterostructure. Li$^+$ ions must diffuse over longer distances to access the MoS$_2$ flake in the heterostructure because the $h$BN flakes completely encase MoS$_2$ (Fig. 2a), such that Li$^+$ ions in the electrolyte do not have direct access to the edges of the MoS$_2$ flake. To investigate this, we fabricated a $h$BN-MoS$_2$-$h$BN heterostructure and reactively ion-etched its perimeter, directly exposing the edges of MoS$_2$ to the liquid electrolyte while still covering the top and bottom surfaces of MoS$_2$ with $h$BN (Fig. 3a). Consistent with the results in Fig. 2, the 2H-1T' phase transition occurred at an applied $V_{EC}$ of 1.0 V vs. Li/Li$^+$ for the edge-exposed heterostructure, as shown by *in situ* Raman spectroscopy and optical microscopy (Fig. 3b and Supplementary Fig. 5). Further intercalation yielded amorphous MoS$_2$ at 0.4 V vs. Li/Li$^+$ just as in the heterostructure without exposed edges. Therefore, despite Li$^+$ ions having direct access to the edges of MoS$_2$, the edge-exposed $h$BN-MoS$_2$-$h$BN heterostructure still required a $V_{EC}$ of 1.0 V vs. Li/Li$^+$ to induce the 2H-1T' phase transition. Thus, longer Li$^+$ diffusion pathways cannot explain the lack of observable difference in V$_{EC}$ required for the phase transition in the heterostructure and bare MoS$_2$.

To further investigate the effects of lithium diffusion pathways, we fabricated a $h$BN-MoS$_2$-$h$BN heterostructure with the heterostructure edges completely covered with a thick layer of gold to block edge-intercalation (Fig. 3c)[32]. With this architecture, Li$^+$ ions could only reach MoS$_2$ through the basal planes of the top $h$BN flake. *In situ* Raman spectra showed the onset of the 2H-1T' phase transition at an applied V$_{EC}$ of 1.0 V vs. Li/Li$^+$ as shown by a mixture of both the A$_{1g}$ and E$_{2g}$ peaks of 2H-MoS$_2$ and the J$_1$ and J$_2$ peaks of 1T'-MoS$_2$ (Fig. 3d and Supplementary Fig. 6). The complete transition to 1T'-MoS$_2$ occurred at 0.8 V vs. Li/Li$^+$, marked by the disappearance of the A$_{1g}$ and E$_{2g}$ peaks. In addition, *in situ* Raman spectra showed MoS$_2$ became amorphous after 15 minutes at a V$_{EC}$ of 0.2 V vs. Li/Li$^+$, instead of at the V$_{EC}$ of 0.4 V vs. Li/Li$^+$ observed in the previous two heterostructures (Fig. 3a-b and Supplementary Figs. 3 and 5). Thus, our results from the edge-covered heterostructure show that, while $h$BN is electrically insulating, it is ionically conductive, allowing Li$^+$ ions to diffuse through its basal planes. We attribute the ionic conductivity of the $h$BN basal plane to intrinsic defects present in the commercially purchased $h$BN crystals (Supplementary Fig. 2d), as pristine 2D $h$BN crystals have been shown to be impermeable even to protons[33], but can become ionically conductive through intrinsic[32,34] or engineered defects[35,36]. The phase transitions in the edge-covered heterostructure were slightly delayed compared to the other two heterostructures, which is consistent with a previous report that suggests an increased energy barrier to lithium diffusion through the basal plane of bare MoS$_2$ as compared to through the interlayer gaps.[32] Irrespective of the Li$^+$ diffusion pathways, all three $h$BN-MoS$_2$-$h$BN heterostructures undergo the 2H-1T' phase transition at 1.0 V vs. Li/Li$^+$, the same V$_{EC}$ to induce the phase transition in bare MoS$_2$. We thus show clearly that the presence of a $h$BN-MoS$_2$ heterointerface does not allow for the 2H-1T' transition at a lower applied V$_{EC}$ than bare MoS$_2$ in contrast to the prediction from DFT calculations.

**Steric hindrance to the phase transition by *h*BN**

Since our calculations indicate that the 1T-MoS$_2$-*h*BN interface is thermodynamically more stable than the 2H-MoS$_2$-*h*BN interface with Li intercalation, the lack of an observable difference in the 2H-1T' phase transition between bare MoS$_2$ and *h*BN-MoS$_2$-*h*BN suggests that there may be an additional energy barrier that influences the phase transition for the heterostructure. We conclude that steric hindrance from *h*BN at the heterointerface introduces an additional energy barrier to the nucleation of 1T'-MoS$_2$ at the interface, counteracting the thermodynamic stabilization of the 1T'-MoS$_2$-*h*BN interface. This is supported by the results from the edge-covered heterostructure (Fig. 3c), where the nucleation of the 1T' phase must be initiated at the top *h*BN-MoS$_2$ interface because the Li$^+$ concentration is highest at the top *h*BN-MoS$_2$ interface as Li$^+$ entered through the basal planes of the top *h*BN flake. The phase transition still occurred at 1.0 V vs. Li/Li$^+$, suggesting an interface-induced barrier to the nucleation of the 1T' phase. We hypothesize that this energy barrier in *h*BN-MoS$_2$-*h*BN heterostructures is driven by sluggish kinetics due to the heterointerface. The phase transition requires atomic rearrangements of the sulfur atoms from a trigonal prismatic to an octahedral coordination surrounding the molybdenum atoms. This rearrangement is likely impeded by steric hindrance induced by the presence of the *h*BN layer in close proximity to the sulfur atoms, thus creating an activation energy barrier for the nucleation of the 1T' phase.

To test our hypothesis that there is an activation energy barrier for the 2H-1T' phase transition introduced by the $h$BN-MoS$_2$ interface, we investigated the phase transition of a $h$BN-MoS$_2$-$h$BN heterostructure as a function of both V$_{EC}$ and temperature (Fig. 4). Increasing the temperature should lower the activation energy barrier and allow for the intercalation-induced phase transition to occur at a higher V$_{EC}$ than the observed 1.0 V vs. Li/Li$^+$ at room temperature. Initially at room temperature, *in situ* Raman spectroscopy showed that lowering V$_{EC}$ to 1.1 V vs. Li/Li$^+$ did not induce the phase change (Fig. 4b), consistent with the other heterostructures. After relaxing the device back to O.C., the microreactor was heated to 60 ºC, and then V$_{EC}$ was lowered to 1.1 V vs. Li/Li$^+$. At 60 ºC, *in situ* Raman indicated the device remained in the 2H phase (Fig. 4c); however, increasing the temperature to 100 ºC for 5 minutes induced the transition to the 1T' phase at 1.1 V vs. Li/Li$^+$, as shown by Raman spectroscopy and optical microscopy (Figs. 4c-d). Thus, the application of thermal energy reduced the applied electrochemical voltage by 0.1 V, suggesting it can overcome the energy barrier to the nucleation of the 1T' phase imposed by the heterointerface. Continued intercalation at 100 ºC for an additional 15 minutes yielded amorphous MoS$_2$. Since the ionic conductivity of the liquid electrolyte at room temperature is already very high (10$^{-3}$ S cm$^{-1}$)[37], we do not expect the application of thermal energy to significantly increase the intercalation of lithium, so the heating experiments further support the hypothesis that the MoS$_2$-$h$BN heterointerface introduces an activation energy barrier to the 2H-1T' phase transition due to steric hindrance.

We studied the role of the heterointerface in the phase stability and transition of a $h$BN-MoS$_2$-$h$BN heterostructure. We show that intercalation-induced phase changes can be manipulated via multi-modal control over applied electrochemical voltage, temperature, and heterointerfaces. While we show that $h$BN-MoS$_2$ interface did not show any observable differences in the electrochemical energy required to induce the MoS$_2$ 2H-1T' phase transition, interfacing MoS$_2$ with other 2D materials may change the kinetics of the phase transition or modify the relative thermodynamic stabilities of 2H- and 1T'-MoS$_2$. This has broad implications for van der Waals heterostructured devices, in which heterointerfaces could modulate the kinetics of phase changes of 2D materials. Given the wide range of phases achievable in 2D materials,[38] our approach suggests opportunities to build devices with targeted electrochemical switching of material properties for a wide variety of applications.

**Methods**

*Ab Initio Calculations:*

The DFT calculations were carried out using the Quantum Espresso[39] package using a plane-wave basis set. Norm-conversing pseudopotentials[40] were used to describe the valence electrons which included the semi-core 4s and 4p electrons of Mo, and the exchange-correlation was treated at the local density approximation[41] (LDA) level, which produces excellent agreement with the experimental structure of bulk 2H and 1T $MoS_2$. The kinetic energy cut-off for the expansion of plane-waves was 2040 eV for all calculations.

We accounted for the effect of doping on the structure of the 2H and 1T $MoS_2$ by relaxing bulk 2H and 1T $MoS_2$ at various levels of implicit doping. The unit cell was allowed to relax until the energy was converged to within 0.02 meV/atom and the forces on atoms were smaller than 2.5 x $10^{-3}$ eV/Å. A Monkhorst-Pack[42] shifted k-mesh of 42x42x8 was used for the calculations. The lattice parameters found in this relaxation were carried over to the calculations on the larger supercell with the heterointerface.

To model the heterointerface, we constructed supercells with one monolayer of 3x3 (2H- or 1T-) $MoS_2$ encapsulated in 6 layers of 4x4 *h*BN (Fig. 1b). We note that while the experiment is performed on few-layer $MoS_2$, the calculations are performed on monolayer $MoS_2$ to reduce computational cost. Increasing the number of $MoS_2$ layers to two does not change our reported results. In the supercell, the lattice parameters of 2H and 1T were constrained to the relaxed parameters of bulk $MoS_2$ (as described above). The stand-alone *h*BN was relaxed in the out-of-plane *c* direction while strained in the *a-b* plane (-4.78 % for 1H and -4.32% for 1T) to create a commensurate heterostructure. Then, the equilibrium distance between $MoS_2$ and *h*BN was determined by minimizing the total energy as the distance between the MoS$_2$ and $h$BN was rigidly adjusted. To determine the binding energy of the MoS$_2$-$h$BN heterointerface, we then calculated the total energy of the supercell with the heterointerface ($E_{hetero.}$), as well as supercells with freestanding monolayer MoS$_2$ ($E_{MoS_2}$) and a 6-layer slab of $h$BN ($E_{BN}^*$, where * shows that the $h$BN was strained as described above) with 25 Å of vacuum. The binding energy is defined as,

$$\bar{E}_{bind} = \frac{E_{hetero.} - (E_{MoS_2} + E_{BN}^*)}{2*A} \qquad (1),$$

where $A$ is the cross-sectional area of the heterointerface. We also tested a larger (less strained) supercell with a 4x4 (2H- or 1T-) MoS$_2$ and 5x5 $h$BN heterostructure consisting of 348 atoms. Since the reduced in-plane strain had little effect on the binding energies of the MoS$_2$-$h$BN interface (Supplementary Table 1), we used the 3x3 MoS$_2$-4x4 $h$BN supercell for the remaining calculations because it is computationally less expensive.

For the Li-doped heterostructures, a Li atom was placed over the Mo site, which is one of the most favored intercalation sites[43], and its separation from the Mo atom was systematically varied to locate the equilibrium position. With a Li atom included, the heterointerface binding energy is accordingly defined as

$$(\bar{E}_{bind,Li}) = \frac{E_{hetero.} - (E_{MoS_2} + E_{BN}^* + E_{Li})}{2*A} \qquad (2),$$

where $E_{Li}$ is energy of an isolated Li atom in vacuum.

*Heterostructure Assembly and Device Fabrication*

MoS$_2$ (SPI Supplies) and *h*BN (HQ Graphene) flakes were mechanically exfoliated from bulk crystals onto SiO$_2$/Si substrates using the scotch-tape method. The substrates were sonicated in acetone and isopropyl alcohol, and treated with O$_2$ plasma prior to exfoliation. Flakes of desired size and thickness were transferred to dry thermal oxide SiO$_2$/Si substrates using a KOH-assisted technique, and MoS$_2$ and *h*BN flakes were stacked on each other to form *h*BN-MoS$_2$-*h*BN heterostructures as follows. A hemispherical droplet of epoxy (Scotch-Weld, Series DP100Plus) about 0.5 mm in diameter was cured on a glass slide. Then, a 13 wt. % solution of polypropylene carbonate (PPC, Sigma Aldrich) in anisole was spun coated onto the epoxy at 3000 RPM for 2 minutes and cured at 90 °C for 2 minutes. Using an optical microscope and a micro-manipulator that holds the glass slide, the epoxy/PPC droplet was positioned above a flake of interest and carefully lowered to contact the flake. About 30 μL of a 2M aqueous solution of potassium hydroxide (KOH, Sigma Aldrich) was added to the substrate to etch the top few Å of SiO$_2$, releasing the flake from the substrate onto the epoxy/PPC droplet (the portion of the glass slide in contact with the KOH was covered with a thin layer of epoxy to prevent etching of the glass). The glass slide with the epoxy/PPC/flake was then rinsed with deionized water to remove any KOH residue. Each flake was released from the glass slide by melting the PPC at 95-100 °C for 5 minutes while contacting a target substrate. The PPC was subsequently dissolved in chloroform overnight.

For electrochemical lithium intercalation, electrodes were patterned with electron beam lithography (Vistec Raith EBPG 5000+) and then 10 nm Cr / 100 nm Au was deposited using thermal evaporation (Mbraun EcoVap). For the heterostructure device with exposed edges, polymethyl methacrylate (PMMA) and hydrogen silsesquioxane (HSQ) were used as a mask for dry etching. First, a layer of PMMA A3 was spun coated on the heterostructure at 4000 RPM for 2 minutes and cured at 180 °C for 2 minutes. HSQ was subsequently spun coated at 3000 RPM with no baking. Electron beam lithography was used to pattern the PMMA/HSQ mask, which was developed in MF-312 developer (Rohm and Haas Electronic Materials) for 4 minutes. After developing, the device was rinsed in multiple DI-water baths in order to remove the developer. The finished mask protected the heterostructure, leaving only the edges exposed for reactive ion etching (Oxford Plasmalab 100). PMMA not covered by HSQ was first removed with $O_2$ plasma (20 sccm) under 50 W RF power for 15 seconds, exposing the edges. A mixture of $O_2$ (4 sccm) and $CHF_3$ (40 sccm) gases under 60 W RF power[44] for 2 minutes was used to etch the edges of the heterostructure. After etching, the PMMA/HSQ mask was removed with acetone.

*Electrochemical Cell Fabrication*

The electrochemical intercalation was performed directly on the heterostructure and bare $MoS_2$ devices. All devices fabricated on $SiO_2$/Si were attached to a glass slide, and the gold contacts of the devices were wire-bonded to copper tape for connection to electrical instrumentation. All subsequent steps were conducted in an argon glovebox.

For all *in situ* Raman experiments, intercalation was conducted with a liquid electrolyte using an enclosed cell that holds the device and electrolyte and is sealed with an optical-grade glass top cover[20,21,32]. Three sides of the glass top cover were first sealed by epoxy, leaving one side open. After the epoxy was cured, a small piece (~3×3 mm) of lithium metal (0.38 mm-thick ribbon, Sigma-Aldrich) was pressed onto copper foil using a mechanical plier to ensure good contact. The lithium/copper foil was then inserted into the open side of the glass top cover. The liquid electrolyte, a battery-grade solution of 1 M lithium hexafluorophosphate in 50/50 v/v ethylene carbonate / diethyl carbonate (LiPF$_6$ in EC/DEC, Sigma Aldrich), was added to the cell to submerge the device and lithium metal. The open side was then covered with epoxy and allowed to cure, forming an air-tight seal.

*In Situ Raman Characterization During Intercalation*

Intercalation cells were connected to a Biological SP300 potentiostat/galvanostat for the electrochemical intercalation of Li$^+$. The Cr/Au contacts to the device served as the working electrode, while the lithium/copper served as the reference/counter electrode. Before intercalation, a Raman spectrum was taken at O.C. (typical O.C. values were 2.4 – 2.7 V vs. Li/Li$^+$). Lithium was intercalated into the heterostructures potentiostatically by dropping $V_{EC}$ vs. Li/Li$^+$ at a scan rate of 10 mV s$^{-1}$. Upon reaching a desired $V_{EC}$, the cell was held at that potential while Raman spectra were collected. For multiple intercalation cycles, cells were allowed to recover to O.C. before the next intercalation. *In situ* heating experiments were conducted by placing the glass slide supporting the intercalation devices onto a hot-plate during intercalation.

All Raman spectra were taken with a Horiba LabRAM HR Evolution Spectrometer using a 633 nm HeNe laser with an 1800 lines/mm diffraction grating. Before intercalation, all samples were characterized at a laser power of ~3 mW to avoid damage, but after cell fabrication, a laser power of ~7.5 mW was used to increase the signal-to-noise ratio due to scattering by the electrolyte. *In situ* Raman spectra were collected with fifteen 5-second exposures.

**Acknowledgements**

S.Y. acknowledges support from the Army Research Office (W911NF-18-1-0367) for device fabrication and intercalation experiments. J.V.P. was supported by the National Defense Science and Engineering Graduate (NDSEG) Fellowship Program, sponsored by the Air Force Research Laboratory (AFRL), the Office of Naval Research (ONR), and the Army Research Office (ARO). Device fabrication and characterization was carried out at the Yale Institute for Nanoscience and Quantum Engineering, the Yale West Campus Materials Characterization Core, the Yale West Campus Cleanroom, and the Yale School of Engineering & Applied Science Cleanroom. The calculations of the heterointerface used resources of the National Energy Research Scientific Computing Center (NERSC), a DOE Office of Science User Facility supported by the Office of Science of the U.S. Department of Energy under Contract No. DE-AC02-05CH11231; the Extreme Science and Engineering Discovery Environment (XSEDE), which is supported by National Science Foundation grant number ACI-1548562; and the Oak Ridge Leadership Computing Facility at the Oak Ridge National Laboratory, which is supported by the Office of Science of the U.S. Department of Energy under Contract No. DE-AC05-00OR22725. We thank the Yale Center for Research Computing use of the research computing infrastructure, specifically the Grace cluster, which was used for calculations on bulk $MoS_2$.

**Author contributions**

S.Y. and J.J.C. conceived the project. S.Y. carried out the experiments and analyzed data with assistance from J.V.P. *Ab initio* calculations were performed by A.K. and D.Y.Q. J.V.P., M.Y., J.M.W., and D.J.H. contributed to the development of experimental methods and characterization techniques. S.Y., J.V.P., and J.J.C. wrote the manuscript with input from all authors.

**Competing interests**

The authors declare no competing interests.

**Additional information**

Supplementary information is available for this manuscript.

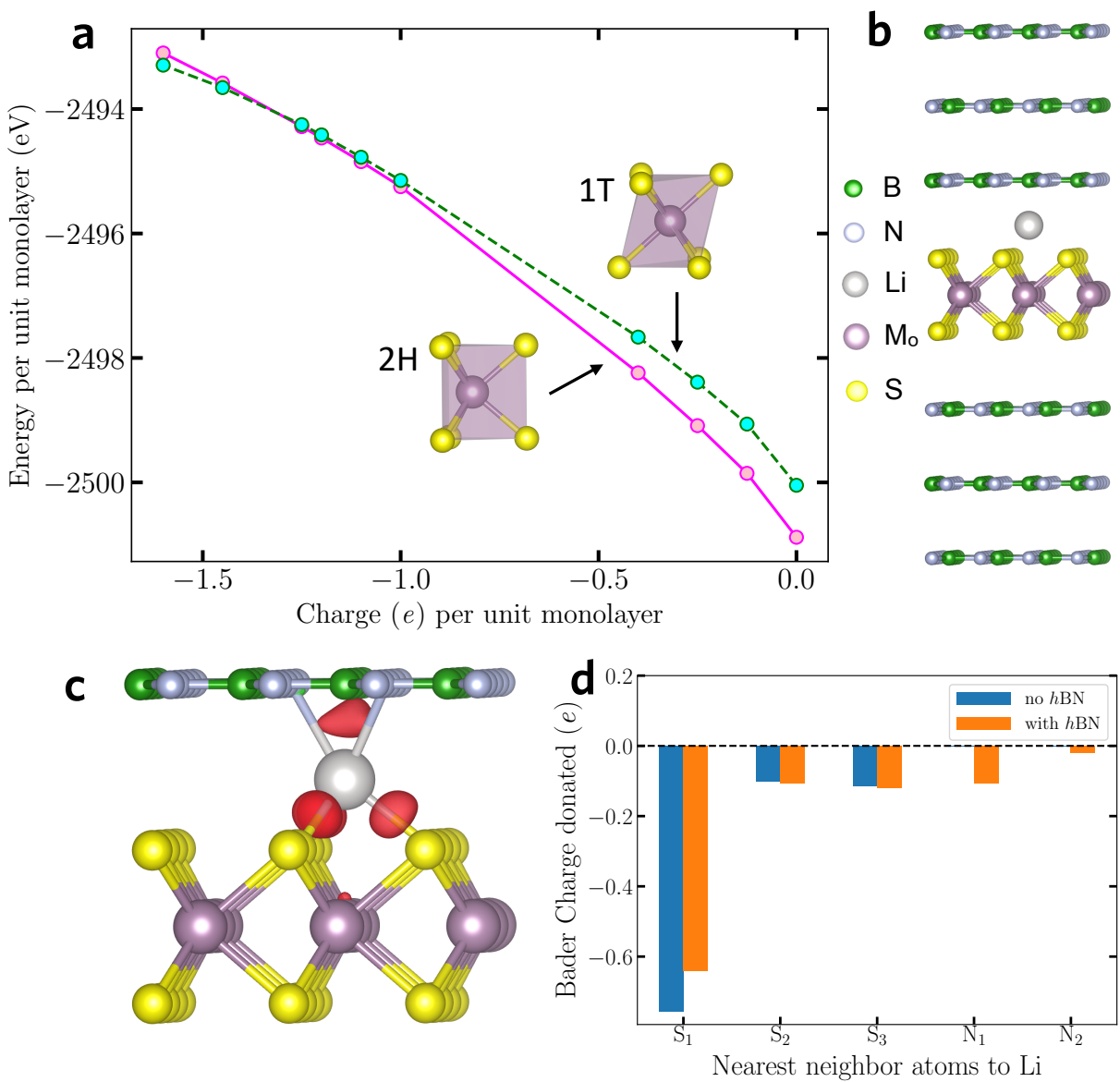

**Fig. 1 | DFT Calculations of the 2H to 1T phase transition. a** Total energy per monolayer of 1 unit cell of bulk $MoS_2$ as a function of added implicit charge for the 2H phase (pink) and the 1T phase (cyan). *Insets*, the trigonal prismatic coordination of sulfur atoms (yellow) around molybdenum (purple) in 2H-$MoS_2$ and their octahedral coordination in 1T-$MoS_2$. **b.** Supercell of the heterointerface with 6 layers of strained $h$BN, 1 layer of relaxed $MoS_2$, and 1 Li atom. **c.** Change in the charge density when Li is added to the heterointerface. Purple atoms are Mo, yellow are S, green are B, light blue are N, and the white atom is Li. Red represents the isosurface corresponding to 80% of the maximum charge density. **d**. Bader charge analysis of the charge donated by the Li atom to all its nearest neighbor S atoms (first nearest - $S_1$, 2$^{nd}$ nearest - $S_2$, and 3$^{rd}$ nearest - $S_3$) and its nearest N atoms ($N_1$ and $N_2$) with (orange) and without (blue) $h$BN encapsulation. The locations of $S_1$, $S_2$, $S_3$, $N_1$, and $N_2$ can be found in Supplementary Fig. 1.

| Heterointerface | $\bar{E}_{bind}$ (mJ/m²) | $\bar{E}_{bind,Li}$ (mJ/m²) |
|---|---|---|
| 2H-MoS$_2$-*h*BN | -139 | -378 |
| 1T-MoS$_2$-*h*BN | -142 | -496 |

**Table 1 | Binding energy of the 2H-MoS$_2$-*h*BN and 1T-MoS$_2$-*h*BN heterointerfaces with and without Li.** The addition of Li increases the binding energy of the 1T-MoS$_2$-*h*BN heterointerface by 354 mJ/m² and the binding energy of the 2H-MoS$_2$-*h*BN interface by 239 mJ/m². Overall, the introduction of Li increases the binding energy of the 1T-MoS$_2$-*h*BN heterointerface relative to the 2H-MoS$_2$-*h*BN heterointerface by 118 mJ/m², whereas the binding energies are almost the same without Li.

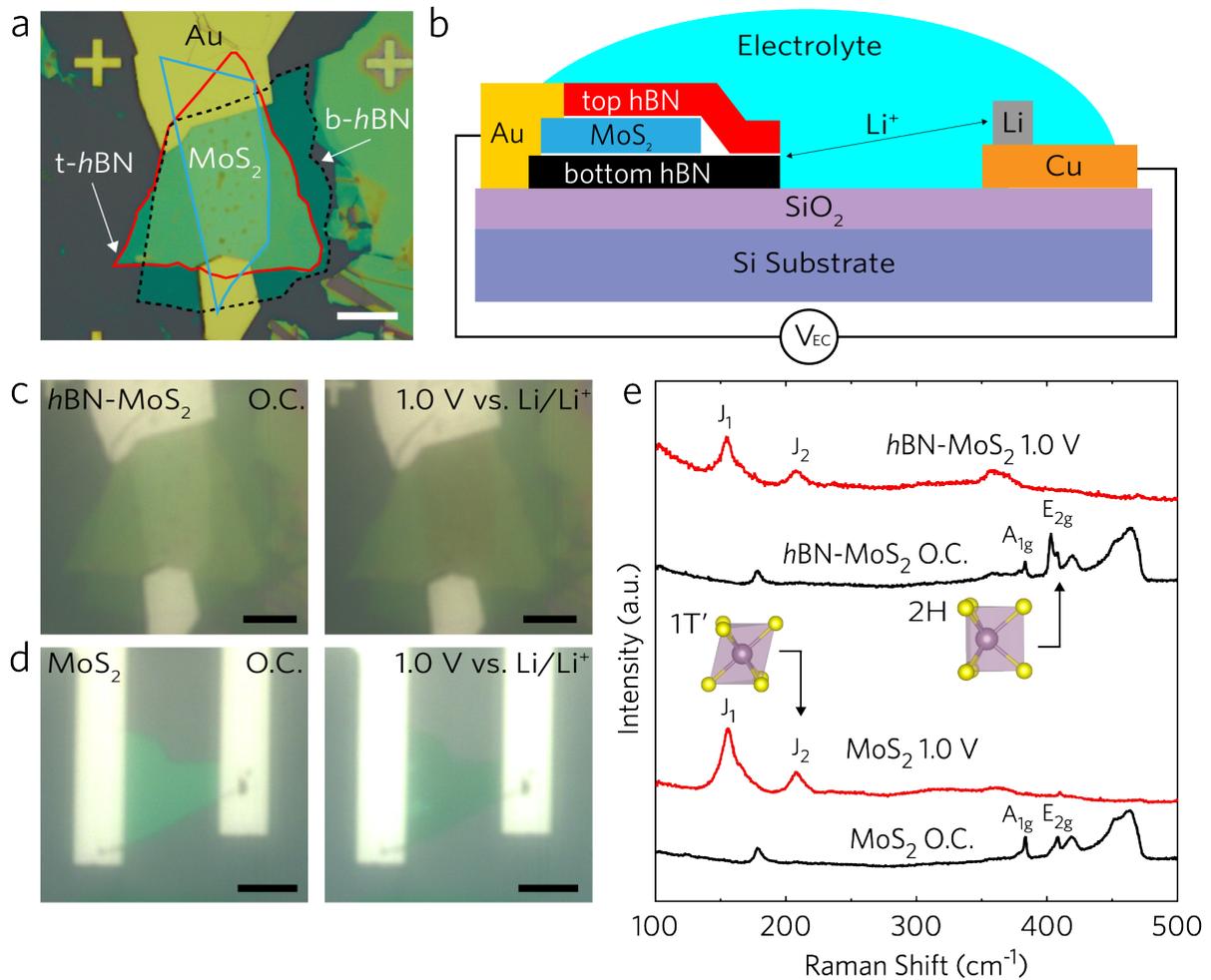

**Fig. 2 | Li$^+$ intercalation of *h*BN-MoS$_2$-*h*BN heterostructures. a,** Optical micrograph of a *h*BN-MoS$_2$-*h*BN heterostructure with Cr/Au contacts to MoS$_2$, scale bar 10 μm. The MoS$_2$ flake (blue outline) is sandwiched between the top and bottom *h*BN flakes, denoted as t-*h*BN (red outline) and b-*h*BN (dashed-black outline), respectively. **b,** Schematic of an electrochemical microreactor for intercalating *h*BN-MoS$_2$-*h*BN. **c,** *In situ* optical micrographs of the heterostructure in (a) at open circuit (O.C.) and 1.0 V vs. Li/Li$^+$, scale bars 10 μm. **d,** *In situ* optical micrographs of bare MoS$_2$ at O.C. and 1.0 V vs. Li/Li$^+$, scale bars 10 μm. **e**, *In situ* Raman spectra of the devices shown in (c-d), with black and red indicating the 2H and 1T' phases, respectively. *Insets*, the trigonal prismatic coordination of sulfur atoms (yellow) around molybdenum (purple) in 2H-MoS$_2$ and their octahedral coordination in 1T'-MoS$_2$.

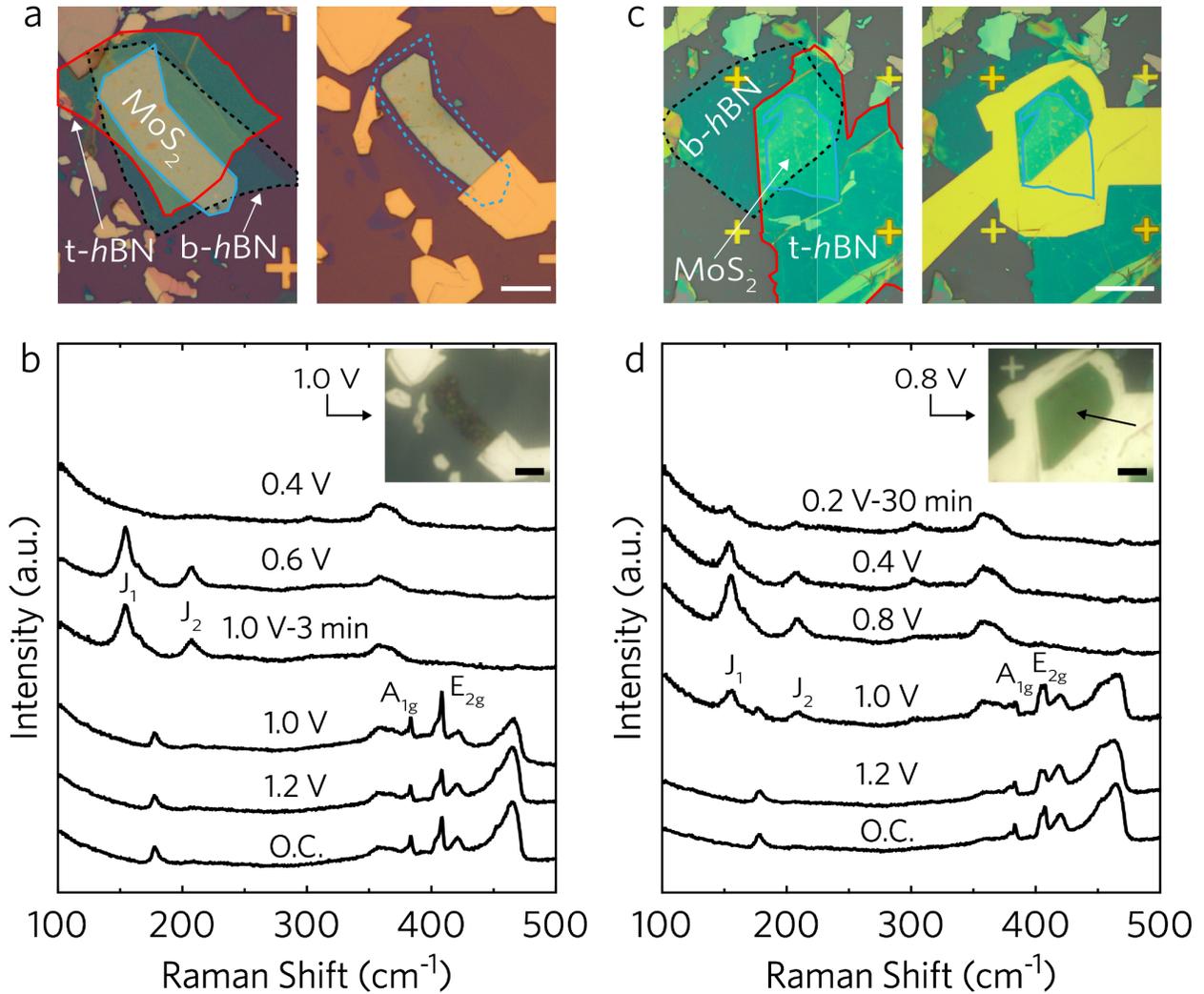

**Fig. 3 | Different Li$^+$ diffusion pathways during intercalation of $h$BN-MoS$_2$-$h$BN heterostructures. a**, Optical micrographs of $h$BN-MoS$_2$-$h$BN before (left) and after (right) etching to expose the edges directly to liquid electrolyte, scale bar 10 μm. The blue-dashed line (right) indicates the original extent of MoS$_2$. **b,** *In situ* Raman spectra of the device in (a) while $V_{EC}$ vs. Li/Li$^+$ is lowered. *Inset*, optical micrograph of the device at 1.0 V vs. Li/Li$^+$, scale bar 10 μm. **c**, Optical micrographs of $h$BN-MoS$_2$-$h$BN before (left) and after (right) covering the edges with gold, scale bar 10 μm. **d,** *In situ* Raman spectra of the device in (c) while $V_{EC}$ vs. Li/Li$^+$ is lowered. *Inset*, optical micrograph of the device at 0.8 V vs. Li/Li$^+$ with a black arrow indicating the darkened region, scale bar 10 μm.

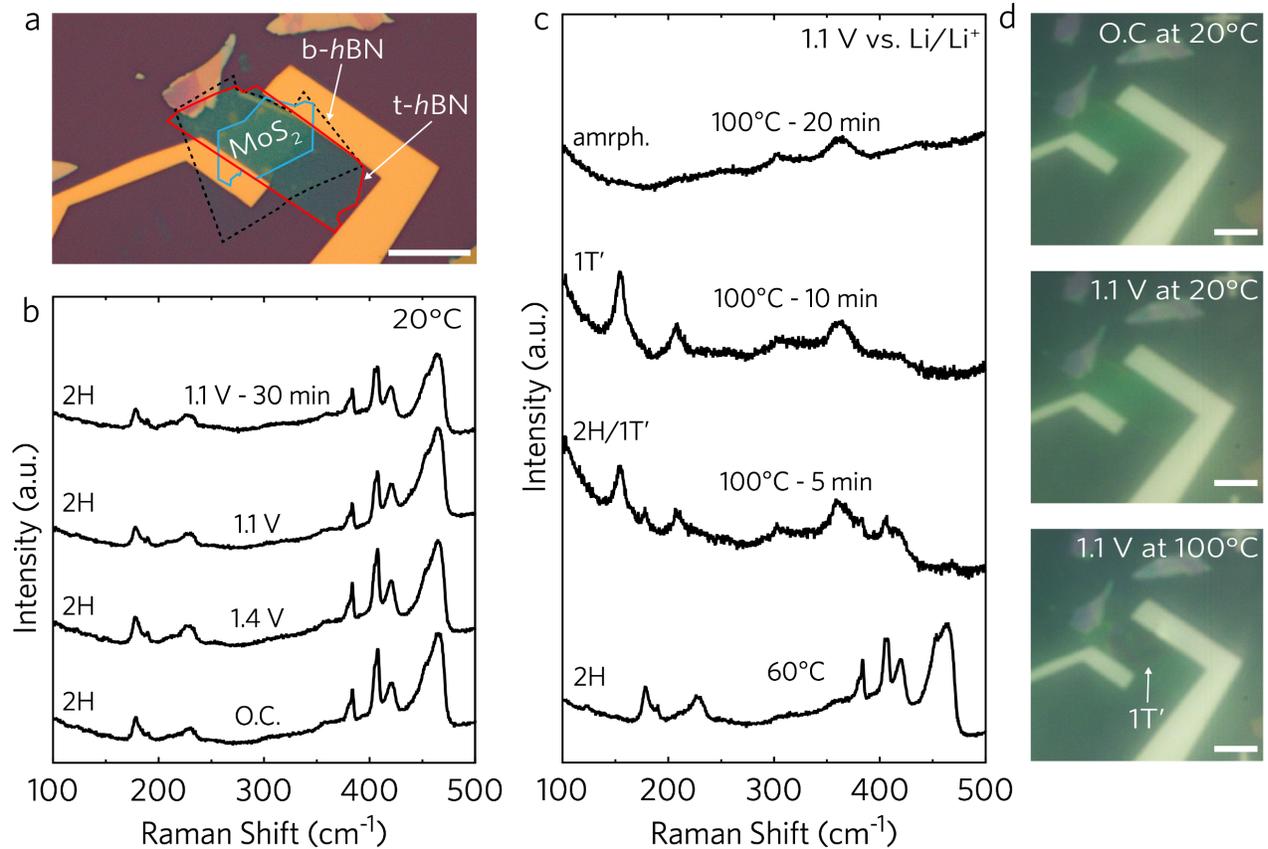

**Fig. 4 | The 2H to 1T' phase transition in *h*BN-MoS₂-*h*BN as a function of temperature. a,** Optical micrograph of the pristine heterostructure device, scale bar 10 μm. **b,** Raman spectra taken as $V_{EC}$ is lowered vs. Li/Li⁺ at room temperature. **c,** Raman spectra taken at 1.1 V vs. Li/Li⁺ during heating. The onset of the 2H-1T' phase transition is observed after 5 minutes at 100°C, with a full transition to 1T'-MoS₂ after 10 minutes. The MoS₂ becomes amorphous after 20 minutes at 100°C. **d,** *In situ* optical micrographs taken at O.C., at 1.1 V vs. Li/Li⁺ at room temperature, and at 1.1 V vs. Li/Li⁺ at 100°C, scale bars 10 μm.